\documentclass[preprint]{aastex61}
\begin{document}

\title{Establishing the Galactic Centre distance using VVV Bulge RR Lyrae variables}

\author{D. Majaess}
\affiliation{Mount Saint Vincent University, Halifax, Nova Scotia, Canada.}
\affiliation{Saint Mary's University, Halifax, Nova Scotia, Canada.}

\author{I. D\'{e}k\'{a}ny}
\affiliation{Astronomisches Rechen-Institut, Zentrum f\"ur Astronomie der Universit\"at
Heidelberg, M\"onchhofstr.~12-14, D-69120 Heidelberg, Germany.}

\author{G. Hajdu}
\affiliation{Astronomisches Rechen-Institut, Zentrum f\"ur Astronomie der Universit\"at
Heidelberg, M\"onchhofstr.~12-14, D-69120 Heidelberg, Germany.}
 
\author{D. Minniti}
\affiliation{Departamento de F\'isica, Facultad de Ciencias Exactas, Universidad Andr\'es Bello, Av.~Fernandez Concha 700, Las Condes, Santiago, Chile.}
\affiliation{Instituto Milenio de Astrof\'isica, Santiago, Chile.}
\affiliation{Vatican Observatory, V00120 Vatican City State, Italy.}

\author{D. Turner}
\affiliation{Saint Mary's University, Halifax, Nova Scotia, Canada.}

\author{W. Gieren}
\affiliation{Instituto Milenio de Astrof\'isica, Santiago, Chile.}
\affiliation{Universidad de Concepci\'{o}n, Departamento de Astronom\'{i}a, Casilla 160-C, Concepci\'{o}n, Chile.}

\begin{abstract}
This study's objective was to exploit infrared VVV (VISTA Variables in the Via Lactea) photometry for high latitude RRab stars to establish an accurate Galactic Centre distance.  RRab candidates were discovered and reaffirmed ($n=4194$) by matching $K_s$ photometry with templates via $\chi^2$ minimization, and contaminants were reduced by ensuring targets adhered to a strict period-amplitude ($\Delta K_s$) trend and passed the Elorietta et al.~classifier.  The distance to the Galactic Centre was determined from a high latitude Bulge subsample ($|b|>4^{o}$, $R_{GC}=8.30 \pm 0.36$ kpc, random uncertainty is relatively negligible), and importantly, the comparatively low color-excess and uncrowded location mitigated uncertainties tied to the extinction law, the magnitude-limited nature of the analysis, and photometric contamination. Circumventing those problems resulted in a key uncertainty being the $M_{K_s}$ relation, which was derived using LMC RRab stars ($M_{K_s}=-(2.66\pm0.06) \log{P}-(1.03\pm0.06)$, $(J-K_s)_0=(0.31\pm0.04) \log{P} + (0.35\pm0.02)$, assuming $\mu_{0,LMC}=18.43$).  The Galactic Centre distance was not corrected for the cone-effect.  Lastly, a new distance indicator emerged as brighter overdensities in the period-magnitude-amplitude diagrams analyzed, which arise from blended RRab and red clump stars.  Blending may thrust faint extragalactic variables into the range of detectability.
\end{abstract}

\section{Introduction}
Distance estimates for the Galactic Centre may be fraught with biases that invalidate otherwise precise evaluations.  The extinction law tied to that sightline is contested and the impact is exacerbated by sizable interstellar reddening. \citet{ma16} confirmed that optical color-excess ratios may vary with stellar population and Galactic position, whereas the near-infrared results are comparatively constant with Galactic longitude ($\ell$) and Galactocentric distance (e.g., $E(J-{3.5\mu m})/E(J-K_s) =1.28\pm0.03$).  \citet{ud03} argued that the optical $R_{V,VI_c}$ extinction law ratio transitions significantly from the Solar neighborhood to the Bulge ($R_{V,VI_c}\sim2.5-2$), and \citet{ca13} advocated that $R_{V,BV}$ is anomalous in Carina.  Near-infrared observations are less sensitive to grain size variations given the implied mean diameter ($< 1.2 \mu m$), although chemistry can play a role at certain wavelengths.  Second, a magnitude-limited bias can be concerning for a lower latitude Bulge sample attenuated by significant extinction.  Mean distances inferred from that sample may shift the Galactic Centre distance toward nearer values.  Blending can magnify that bias, and may be enhanced near the crowded Galactic Centre.  \citet{sm07} remarked that effect can impact red clump determinations of the Galactic Centre distance and attempts to deduce parameters for the putative bar(s), and blending can be acute within the galaxies and globular clusters used to constrain cosmological models \citep[e.g.,][]{ma12b}.  High resolution HST data imply that multiple stars are in close proximity to RR Lyrae stars near the core, where the stellar density increases markedly, and numerous neighbouring stars remain unresolved in ground-based images.  

The aforementioned uncertainties are mitigated in the present analysis owing to the near-infrared nature of the VVV (VISTA Variables in the Via Lactea) survey data utilized, in tandem with the selection of uncrowded high latitude Bulge RR Lyrae stars exhibiting relatively marginal reddening.  The discovery  of such stars in the VVV survey is important given those objectives, in addition to standard candles generally being employed to characterize interstellar extinction, identify stellar streams and globular clusters, and facilitate efforts to delineate the structure of the Milky Way \citep[e.g.,][]{de15,ma17}. 

In this study, a search is performed for higher latitude Bulge RRab stars in the VVV survey with the aim of establishing an accurate Galactic Center distance.  An emphasis is placed on low extinction and relatively uncrowded stars at higher latitudes, thereby reducing the uncertainties.   This work is organized as follows: \S \ref{sec-vvv} and \S \ref{sec-rrab} describe the VVV survey and the method employed to detect RR Lyrae stars therein; in \S \ref{sec-gc} a subsample of the candidates identified are utilized to ascertain the distance to the Galactic Centre, and the zeropoint is tied to LMC variables; \S \ref{sec-nd} features a brief characterization of a potential new distance indicator present in the data analyzed; and concluding remarks are provided in \S \ref{sec-conclusion}.    

\begin{figure}[!t]
\begin{center}
\includegraphics[width=8.5cm]{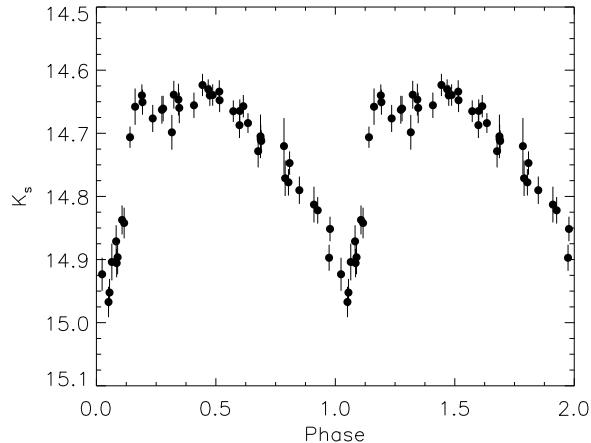} 
\caption{VVV RRab candidates were identified using the GLS period-search algorithm \citep{zk09} and a suite of criteria (e.g., the \citealt{el16} classifier).  A high latitude candidate ($\ell,b \approx 3,-10^{\circ}$) featuring low reddening is displayed (phase mirrored to elucidate the lightcurve).}
\label{fig-lc}
\end{center}
\end{figure}

\section{Analysis}
\subsection{VVV Survey}
\label{sec-vvv}
The VVV survey is a near-infrared, wide-field, multi-epoch campaign monitoring stars in the Galactic Bulge, and adjacent regions of the disk \citep{mi10,sa12}.  Preliminary analyses of the VVV data have supported important conclusions related to numerous topics, and a brief discussion is warranted.  \citet{ma16} relied partly on VVV RR Lyrae variables to investigate properties of dust extinction, which remains a critical topic in the forthcoming Gaia era as parallaxes require conversion to absolute magnitude space, thus ensuring the gains can be transferred to the extragalatic scale and beyond Gaia's limits.   \citet{de13} and \citet{mi17a} advocated that Bulge RR Lyrae variables sampled by the VVV campaign do not trace the bar(s) outlined by their younger red clump counterparts.\footnote{For an alternative interpretation see \citet{co06} and \citet{pi15}.}  Furthermore, \citet{gr16} concluded that the spatial distribution of VVV RR Lyrae stars differs relative to the X-shaped structure demarcated by red clump stars.  \citet{mi17b} discovered an Oosterhoff type I globular cluster (FSR 1716) using RR Lyrae variables identified in the VVV survey, and the effort is part of a broader objective to help rectify an incomplete census of Galactic globular clusters.   

The current analysis relies on preliminary VVVDR4 aperture photometry produced by the VISTA Data Flow System (VDFS) pipeline \citep[][and references therein]{cr12}.  The vvvDetection, vvvVariability, and vvvSource tables were queried.  All VDFS VVV photometry and catalogs will be inevitably released to the public.  

\subsection{Identifying RRab Stars}
\label{sec-rrab}
\begin{figure}[!t]
\begin{center}
\includegraphics[width=8.5cm]{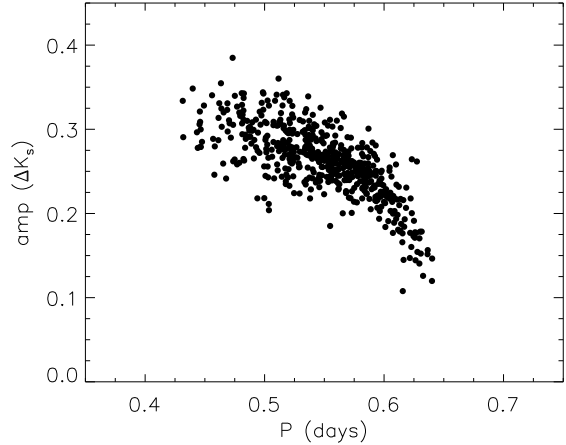} 
\caption{Contamination from other variables was reduced by relying partly upon the RRab period-amplitude ($\Delta K_s$) diagram and the \citet{el16} machine-learned classifier.}
\label{fig-ksamp}
\end{center}
\end{figure}

A suite of criteria were adopted to avoid stars that mimic RRab characteristics.  For example, phased RRab $K_s$ lightcurves may exhibit symmetry owing to the reduced sensitivity to temperature (i.e., Rayleigh-Jeans tail of the Planck function), and can therefore display similar lightcurves to shorter-period contact eclipsing binaries.  A period search for RRab candidates was carried out on the VVVDR4 data using the GLS algorithm \citep{zk09}, which produces a generalised Lomb-Scargle periodogram.  The periods were restricted to $0.41\leq P \leq 0.65$ days and a power signal strength $>0.5$, whereby the former limit aimed to mitigate contamination by other variables, as the bulk of the RRab distribution lies within that range \citep[][their Fig.~1]{so09}.  Second, $K_s$ template lightcurves inferred from OGLE data \citep{so14} were fitted to the targets using $\chi ^2$ minimization.  The candidates exhibited multiband $JHK_s$ colours encompassing the O-M stellar reddening line and adhered to the trend displayed in the $K_s$ period-amplitude diagram (Fig.~\ref{fig-ksamp}): $\Delta K_s \approx -9.65 P^3 + 10.72 P^2 - 3.94 P + 0.803$.  That polynomial was approximated using an OGLE subsample, and an amplitude deviation threshold of $\approx 0.08$ was adopted. The amplitude criterion may preferentially select RRab of a particular chemical composition, however, near-infrared distances are comparatively insensitive to metallicity \citep[e.g.,][]{mu15}.  Importantly, only candidates that were subsequently deemed as RRab stars by the \citet{el16} machine-learned classifier are analyzed here ($\approx 1000$ targets eliminated with a mean near 10.5 kpc).  The objective being to provide an independent check on the classifications and reduce contamination, rather than strive for completeness. 

The approach yielded $4194$ candidate RRab stars, with $2849$ targets in common with \citet{gr16} and OGLE \citep{so14}, thereby bolstering the candidate classifications issued by all three studies owing partly to the strict period-$\Delta K_s$ constraint.   The objects typically feature 50 $K_s$ datapoints, and span a sizable Galactic longitude and latitude baseline ($350 \lesssim \ell \lesssim 10^{o}$, $-10 \lesssim b \lesssim -2.7^o$, $2.7 \lesssim b \lesssim 5^o$).  A digitized table of the data (e.g., number of epochs, etc.) will accompany this work and shall be hosted by CDS.  Certain DR4 regions feature additional epochs and depth, thereby fostering potential biases.  Those issues shall be ameliorated by the forthcoming DR5 release and the desirable VVV\textit{x} (extended) survey, and VVV sourceIDs for the targets will be cited within the table to facilitate the acquisition of the latest data.  

\subsection{Galactic Centre Distance}
\label{sec-gc}
\begin{figure}[!t]
\begin{center}
\includegraphics[width=8.5cm]{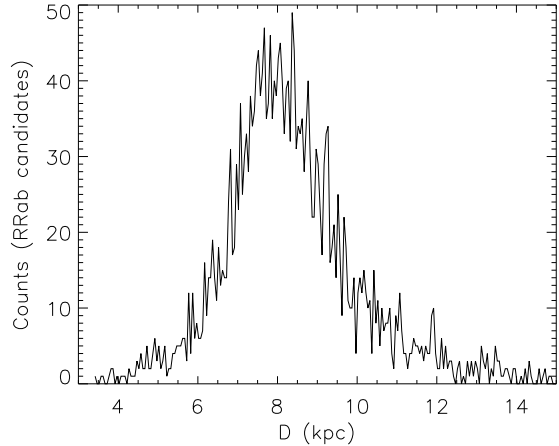} 
\caption{High latitude ($|b|>4^{o}$) and low extinction RRab candidates were used to infer the Galactic Centre distance, thereby mitigating numerous biases. The mean distance inferred from simulations (10000) where the coefficients and parameters (extinction law, histogram binsize, etc.) were varied is $R_{GC}=8.30\pm0.36$ kpc.  The histogram shown is an example iteration tied to optimal coefficients (peak near $8.4$ kpc).}
\label{fig-hi}
\end{center}
\end{figure}

The Galactic Centre distance was computed using near-infrared photometry for a higher latitude subsample of the RR Lyrae stars identified ($|b|>4^{o}$).  A mean distance of $R_{GC}=8.30 \pm 0.36$ kpc was inferred from simulations (n=10000) where coefficients and parameters (period-magnitude, extinction law, histogram binsize, etc.) were varied. A single example iteration is conveyed as Fig.~\ref{fig-hi}, and the distance tied to the distribution's maximum was adopted.  The distance derived agrees with other recent estimates \citep{tu14,db16,bh17}, including \citet{bg16}, who obtained $8.2\pm0.1$ kpc.   \citet{ma10} discussed the uncertainty in characterizing how the mean distance to a group of variable stars relates to $R_{GC}$, however, solace is taken in the convergence of several semi-independent VVV estimates.  The distances, heliocentric and projected onto the Galactic plane, were evaluated via the relations derived below ($M_{K_s}$, $(J-K_s)_0$).  Distances for a lower latitude subsample were expectedly closer owing to increased extinction and photometric contamination.  Brighter candidates detected by 2MASS exhibit an even nearer distance, reiterating the prominent role $\chi^2$ minimization and extinction play in shifting the results. The example histogram's asymmetry at smaller distances emphasizes those points, which dominate over the reputed cone-effect in this instance \citep[see][and discussion therein]{de13,bh17}. The distance was not corrected for that effect. Finally, the selection of a higher latitude sample is likewise beneficial because the VVV-2MASS standardization is complicated at lower latitudes where extinction is highest \citep{gf17}, and the VDFS VVV data used were extracted by aperture photometry which is sensitive to crowding inherent to such regions.

\begin{figure}[!t]
\begin{center}
\includegraphics[width=8.5cm]{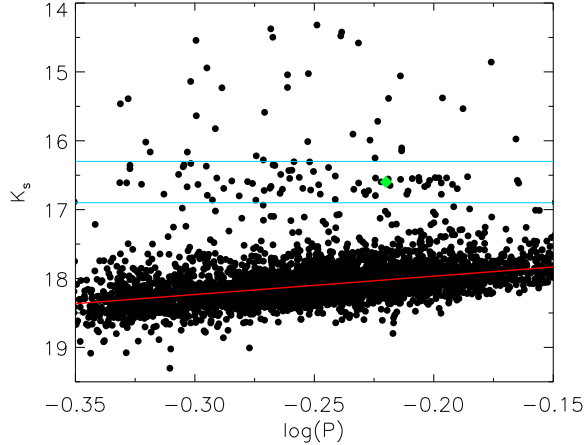} 
\caption{The period-$K_s$ trend (red line) derived from LMC OGLE RRab stars with VMCDR4 photometry.  The absolute magnitude function implied by iterative clipping is: $M_{K_s}=-2.66 \log{P} + (17.40-\mu_{0,LMC})$.  Brighter stars were culled prior to the fitting process, including the overdensity (cyan lines) representing single blended RRab and red clump stars (green diamond, \S \ref{sec-nd}).}
\label{fig-pm}
\end{center}
\end{figure}

The distances were evaluated using $M_{K_s}$ and $(J-K_s)_0$ relations inferred from LMC stars.  The strategy employed to reduce the biases described above ensured that key uncertainties stem partly from those functions.  \citet{mu15} summarized the ambiguity surrounding the $M_{K_s}$ relation in their Table~3.  The topic was therefore investigated using $\approx 4500$ LMC RRab identified by OGLE and possessing VMCDR4 data \citep{so09,ci11}.  The VISTA VMC data, in concert with the VISTA VVV data, will invariably be publicly available.  Means were adopted verbatim from the preliminary VDFS VMCDR4 source table (aperture photometry), and are used in the absence of the final data release when multiband template fitting to PSF photometry will be performed (Hajdu et al., in prep.).  The coefficients determined from that sample support  the \citet{mu15} findings and are: $M_{K_s}=(-2.66\pm0.06) \log{P} + (-1.03\pm0.06)$.  The relation was established by first applying a robust fit, removing $3 \sigma$ outliers, and subsequently taking the mean of robust and least-squares algorithms (Fig.~\ref{fig-pm}).   Brighter blends between RRab and red clump stars were excluded from the fitting process (\S \ref{sec-nd}).  Reddenings associated with the Bulge RRab distances were computed via the following expression derived from the VMC sample: $(J-K_s)_0=(0.31\pm0.04) \log{P} + (0.35\pm0.02)$.  An LMC reddening and distance modulus of $E(B-V)=0.13\pm0.02$ and $\mu_0=18.43\pm0.03$ were assumed \citep{ma10,ma11b}, accordingly.  The following color-excess ratio and extinction law of $E(J-K_s)/E(B-V)=0.48\pm0.01$ and $A_{K_{s}}/E(J-K_s)=0.49\pm0.02$ \citep{ma16} were likewise adopted, respectively.  The former was evaluated by correlating optical and near-infrared photometry tabulated for Galactic Cepheids by \citet{be00} and \citet{mp11}, and drawing upon the expressions conveyed by \citet{ma16}, and the latter was selected owing to the reduced extinction at $K_s$ and the ratio is corroborated by \citet{mi18} who obtained $A_{K_{s}}/E(J-K_s)=0.484\pm0.040$ \citep[see also][]{al17}.  Formal fitting uncertainties are cited and ignore the metallicity sensitivity of the $(B-V)$ color index \citep{ma16}.  However, overall, the resulting corrections in the infrared domain are advantageously small and the uncertainties comparatively negligible.   The uncertainty for the $M_{K_s}$ zeropoint was enlarged to account for differences between the \citet{ma11b} and Auracaria-OGLE distances to the LMC \citep{pi13,gi15}.  The latter obtained $\mu_0=18.497$, which shifts the zeropoint to $-1.09$.  Those interested in adopting an alternative LMC distance may use:  $M_{K_s}=-2.66 \log{P} + (17.40-\mu_{0,LMC})$.   Note that the relations conveyed are on the 2MASS system, and the VVV-VMC DR4 data were transformed from the VISTA system using common stars.

The analysis could be admittedly improved by enhanced determinations of the mean magnitudes from multiband template fitting to PSF photometry (Hajdu et al., in prep.).  The preliminary DR4 aperture photometry utilized here would benefit from that approach, which shall bolster the reddenings, distances, and the photometry's accuracy.  Shorter-wavelength DR4 VISTA YZJH observations also possess few epochs relative to $K_s$ (e.g., one $J$ sampling in VVVDR4), and that shall be expanded in subsequent releases.

\subsection{Blended Distance Indicator}
\label{sec-nd}
\begin{figure*}[!t]
\begin{center}
\includegraphics[width=11.5cm]{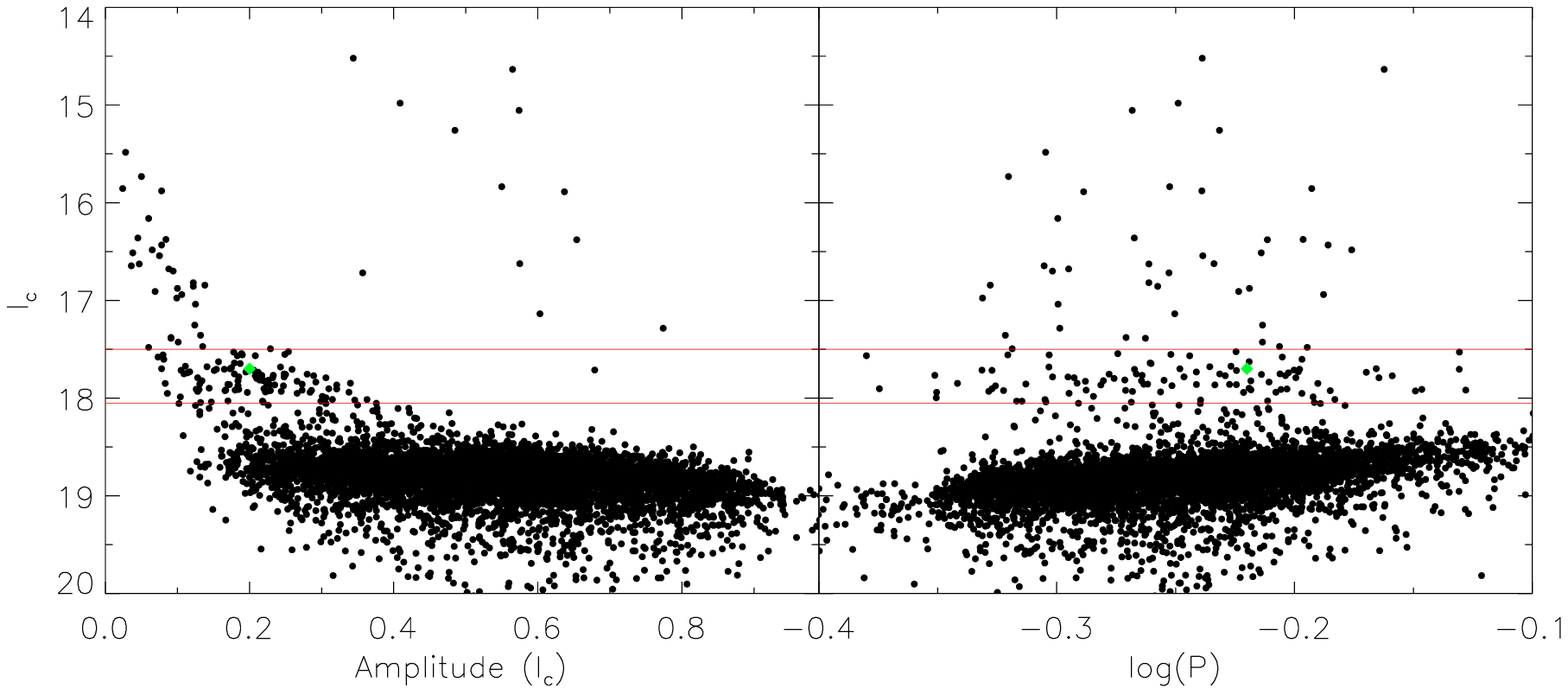} 
\includegraphics[width=6.4cm]{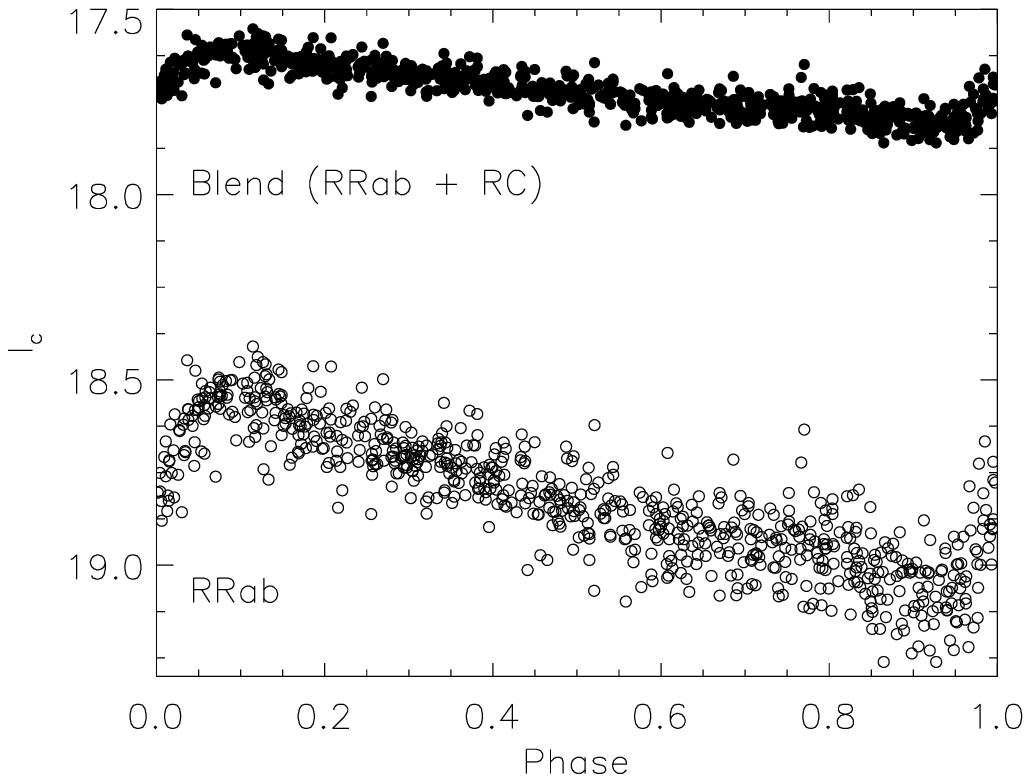} 
\caption{Blended LMC RRab and red clump stars exhibit a smaller amplitude and increased brightness (right), and the result (green diamond) is an overdensity displayed in magnitude diagrams (left, see also Fig.~\ref{fig-pm}).  Blending may propagate fainter extragalactic RRab stars into the detection range.}
\label{fig-bl}
\end{center}
\end{figure*}

The period-magnitude diagram exhibited an overdensity $\approx 1^{m}$ brighter than canonical RR Lyrae stars (Fig.~\ref{fig-pm}).  OGLE data were subsequently examined in amplitude-magnitude space, and the overdensities are associated with brighter targets mimicking smaller-amplitude RRab lightcurves (Fig.~\ref{fig-bl}).  The objects are forged from blends between RRab stars and the common field red clump giant.  Advantages of red clump stars include their sizable statistics and ubiquity.  \citet{al02} deduced apparent magnitudes of $K_s=16.97$ and $I_c=18.21$ for red clump stars in the LMC, which when paired with brightness estimates for a canonical RRab match the observed overdensity locations (green diamond, Figs.~\ref{fig-pm}, \ref{fig-bl}).  Fig.~\ref{fig-bl} displays a $P\sim 0.6^{\rm d}$ RRab impacted by blending, and a diminished amplitude is noticeable as a result.  The overdensity could be a distance indicator governed by a mean or maximum (edge detection) magnitude threshold, in a fashion reminiscent of red clump or TRGB analyses.  A viable methodology may emerge with subsequent iterations and independent input, with the frequency (not location) of single RRab and red clump blends being partly dictated by instrumentation and seeing.  Furthermore, the period-amplitude diagram (Fig.~\ref{fig-bl}) may contain blends constituted by multiple stars, and thus modelling could proceed on an individual target basis rather than overdensities.  Amplitude and brightness ratios in multiple passbands can constrain the quantity of red clump stars involved.  At remote distances blending may thrust RR Lyrae stars and other variables beyond faint-limit thresholds.  A subsequent work will explore whether the approach can be utilized to establish relative distances between the Milky Way, LMC, SMC, IC1613, and comparatively nearby galaxies.  Relative distances can be employed to constrain the impact of metallicity (e.g., [$\rm Fe/H$]) for separate standard candles \citep[e.g., classical Cepheids,][]{ma11}.

\section{Conclusion}
\label{sec-conclusion}
In this study near-infrared VVV photometry for higher latitude RRab stars were employed to determine the Galactic Centre distance, thus minimizing numerous uncertainties. 

RRab candidates (e.g., Fig.~\ref{fig-lc}) were classified by matching $K_s$ photometry to templates via $\chi^2$ minimization, and contaminants were mitigated by relying on the period-$\Delta K_s$ diagram and a machine-learned classifier \citep{el16}. 4194 RRab candidates were identified, with $2849$ targets in common with \citet{gr16} and OGLE, thus strengthening the candidate classifications of all three research initiatives owing in part to the inclusion of a conservative near-infrared period-amplitude criterion (Fig.~\ref{fig-ksamp}). 

A low extinction and high latitude ($|b|>4^{o}$) subsample was exploited to determine the distance to the Galactic Centre ($R_{GC}=8.30 \pm 0.36$ kpc, Fig.~\ref{fig-hi}).  The candidates were selected to reduce the impact of certain biases, granted sampling high extinction targets in crowded regions will yield nearer distances owing to the magnitude-limited nature of the analysis, the $\chi^2$ minimization adopted to identify the variable class, and photometric contamination.  The computed distance to the Galactic Centre is linked to $M_{K_s}$ and $(J-K_s)_0$ relations inferred from LMC RR Lyrae variables ($M_{K_s}=-(2.66\pm0.06) \log{P}-(1.03\pm0.06)$, $(J-K_s)_0=(0.31\pm0.04) \log{P} + (0.35\pm0.02)$).  Overdensities were identified and culled from the LMC data used to establish the period-magnitude relation (Fig.~\ref{fig-pm}), and arise from blends of RRab and red clump giants (Fig.~\ref{fig-bl}).  The objects could be a viable distance indicator owing to the enhanced brightness and lightcurve morphology.

Lastly, with subsequent iterations the final VVV catalog, in tandem with other surveys (e.g., OGLE), may continue to deliver pertinent information regarding the distance, reddening, age, and metallicity of stellar populations throughout the Galaxy's central region \citep[e.g.,][]{mi18,de18}.  
 
\acknowledgments
\footnotesize{Acknowledgements: D.M.~(Majaess) is grateful to the following individuals and consortia whose efforts, advice, or encouragement enabled the research: OGLE, VSA (N.~Cross \& M.~Read), CDS, 2MASS, arXiv, and NASA ADS.  I.D.~was supported by Sonderforschungsbereich SFB 881 ``The Milky Way System'' (subproject A3) of the German Research Foundation (DFG).  Data analysis was partly carried out on the Milky Way supercomputer, which is funded by the Deutsche Forschungsgemeinschaft (DFG) through the Collaborative Research Center (SFB 881) ``The Milky Way System'' (subproject Z2).  D.M.~(Minniti) and W.G.~gratefully acknowledge support from the Millenium Institute of Astrophysics (MAS) of the Iniciativa Cientifica Milenio del Ministerio de Economia, Fomento y Turismo de Chile, grant IC120009, and the BASAL Centro de Astrofisica y Tecnologias Afines (CATA) PFB-06/2007.  D.M.~(Minniti) likewise acknowledges support from FONDECYT Regular grant No.~1170121.  W.G.~also acknowledges support from the European Research Council (ERC) under the European Unions Horizon 2020 research and innovation program (grant agreement No.~695099).}

\end{document}